\documentclass[preprintnumbers,amsmath,amssymb,floatfix,14pt,prd,ocolumn,showpacs,
superscriptaddress,nofootinbib]{revtex4}
\usepackage[colorlinks=true, pdfstartview=FitV, linkcolor=blue, citecolor=red, urlcolor=magenta, breaklinks=true]{hyperref}
\usepackage[dvips]{graphicx}

\usepackage{epsfig}
\usepackage{graphicx}
\usepackage{indentfirst}
\usepackage{amsmath}
\usepackage{amsfonts}
\usepackage{amssymb}
\usepackage{array}
\begin{document}
\title{Geodesic Deviation Equation in $\Lambda$CDM  $f(T,\mathcal{T})$ gravity }
\author{M. G. Ganiou}
\email{moussiliou_ganiou @yahoo.fr}
\affiliation{Institut de Math\'ematiques et de Sciences Physiques
(IMSP), Universit\'e de Porto-Novo, 01 BP 613 Porto-Novo, B\'enin}
\author{Ines G. Salako}
\email{inessalako@gmail.com}
\affiliation{Institut de Math\'ematiques et de Sciences Physiques
(IMSP), Universit\'e de Porto-Novo, 01 BP 613 Porto-Novo, B\'enin}
\affiliation{D\'epartement de Physique ,  Universit\'e d'Agriculture
de K\'etou, BP 13 K\'etou,  B\'enin}
\author{ M. J. S. Houndjo}
\email{ sthoundjo@yahoo.fr}
\affiliation{Institut de Math\'ematiques et de Sciences Physiques
(IMSP), Universit\'e de Porto-Novo, 01 BP 613 Porto-Novo, B\'enin}
\affiliation{Facult\'e des Sciences et Techniques de Natitingou (FAST), Universit\'e de Natitingou, BP 72 Natitingou, B\'enin}
\author{J. Tossa }
\email{joel.tossa@imsp-uac.org}
\affiliation{Institut de Math\'ematiques et de Sciences Physiques
(IMSP), Universit\'e de Porto-Novo, 01 BP 613 Porto-Novo, B\'enin}

\begin{abstract}
\hspace{0,2cm} 
The geodesic deviation equation has been investigated in the framework of $f(T,\mathcal{T})$ gravity, where $T$ denotes the torsion and $\mathcal{T}$
is the trace of the  energy-momentum tensor, respectively. The FRW metric is assumed and the geodesic deviation  equation has been established
following the General Relativity  approach in the first hand and secondly, by a direct method using the modified Friedmann equations. Via 
fundamental observers and null vector fields with FRW background,  we have generalized the Raychaudhuri equation and the Mattig relation in 
$f(T,\mathcal{T})$ gravity. Furthermore,  we have numerically solved  the geodesic deviation equation for null vector fields by considering 
a particular  form  of  $f(T,\mathcal{T})$ which induces  interesting results susceptible to be tested   with observational data.

\end{abstract}
\maketitle
\pretolerance10000

\tableofcontents

\section{Introduction}
General Relativity (GR), still known as the general theory of relativity is a geometrical theory of gravitation
  published by Albert Einstein in 1915 and also a current description of gravitation in modern physics.
  Relativity General generalizes special relativity and Newton's laws of gravitation and thus provides an unified description of 
  gravity as a rise property of the geometry of space and time, or at least space-time. In particular,
  the curvature of space-time is directly related to the energy and momentum generated by presence of matter and radiation in space-time.
  This interaction between the geometry and the energy-momentum is specified by Einstein's equations, a system of  partial differential equations.
  One of the most features studied in this successful theory of Einstein (GR) is the Geodesic Deviation Equation (GDE)  \cite{Misner}. Indeed,
  the relative motion  of test particles is one of the important way to get informations on the gravitational field and on the  space-time geometry.
  This motion  is described by the Geodesic Deviation Equation . It can be claimed that the GDE is also one
  of the most important equations in relativity for the fact that it gives a way to step the curvature of space-time.
  This feature in relativity has been discussed by Szekeres \cite{eti25}. The curvature of space-time described by Riemann
tensor manifests  through the GDE \cite{eti20}-\cite{eti22} and the relative  acceleration of test particules also known under the denomination 
tidal acceleration.  The meaning of GDE in relation with the relative acceleration and tidal forces between two neighboring particles in free 
fall under the influence of gravity was  so stressed for several times by  half of the fifties.  
  The importance of GDE for spinless particles was emerged by the authors \cite{eti23},\cite{eti24}. They have notified this 
  importance when they study  the gravitational waves and their detection.  The GDE provides a graceful tool
  to explore  time-like, space-like and null structures of space-time  geometry  and we can get, by solving it, the following important relations:
  the   Raychaudhuri equation \cite{eti26}, the Mattig relation  \cite{eti27} and the Pirani equation \cite{eti28}.\par 
  Actually, the current accelerating Universe was strongly confirmed by several independent experiments such as
Radiation of Cosmic Microwave Background (CMBR) \cite{Spergel} and the Sloan Digital Sky Survey (SDSS) \cite{Adelman}. This state of Universe is 
attempted to be explained in the litterature  by two approaches. The first approach involves the correction of General Relativity and  the
second assumes that the Universe is dominated by an antigravity component  called dark energy.  All these approaches are based on the fundamental
theories of gravity
namely General Relativity and Tele-parallel Theory equivalent of GR
(TEGR)\cite{a'} and on  the  modified gravity theories. Severals  modified theories  have been constructed by modification of GR:( $f(R)$, $f(R,\mathcal{T})$ \cite{ma1}-\cite{ma6}, 
$f(G)$ \cite{mj1}-\cite{mj5}) where $R$ is the scalar curvature, $\mathcal{T}$ is the trace of energy-momentum tensor and $G$ the Gauss-Bonnet invariant
defined by $G=R^2-4R_{\mu\nu}R^{\mu\nu}+ R_{\mu\nu \lambda\sigma}R^{\mu\nu\lambda\sigma}$. Furthermore, an another approach
which in spirit, is similar to the first modification of GR is the so called $f(T)$ theory which is 
the modified version of TEGR with  $T$ the scalar torsion. \\
Indeed, instead of the connection of Levi-Civi as it is done in GR, $f (T)$  gravity  uses the Weitzenbok connection. This theory was introduced for the first time by  Ferraro\cite{Ferraro1} in  their work on  
UV modification of GR and  inflation. Very soon after and  in the context of modern cosmology,
Ferraro and Bengochea \cite{Ferraro2} considered the same theory to describe dark energy. Other studies can be cited
as examples: \cite{st1}- \cite{st41}. Here, a special attention was tuned to the theory  $ f(T, \mathcal {T}) $ as an extension of the
 $ f (T) $ gravity  with $ \mathcal {T} $ the trace of energy-momentum tensor. Severals works with interessing results have been developped in this 
 framework  \cite{sala1}- \cite{sala4}.\par 
  The GDE has already been studied in  $f(R)$ theory  \cite{eti35},\cite{eti36}, $f(T)$ theory and also $f(R,\mathcal{T})$ gravity \cite{etienne} 
  where $ R $, $ T $ and $ \mathcal {T} $ are still the objects  defined earlier. Excellent results have
  been found  with these above theories. Our main goal in this present work is to find a new approach
   to get the GDE in the $ f (T, \mathcal {T}) $  gravity. \par 
    This paper is organised as follow: In Sec \ref{sec2} we introduce the gravity $ f (T, \mathcal {T}) $ 
    and obtain its field equation in cosmology FLRW . The Sec \ref{sec3} reminds the GDE in the context of General Relativity and we have 
    extended this study to  
    $ f(T, \mathcal {T}) $ theory at Sec  \ref{sec4}. A conclusion comes naturally to sanction the end of our work
  in Sec  \ref{sec5}.

 \section{Generality on $f(T,\mathcal{T})$  gravity within  FLRW Cosmology}\label{sec2}
 In Tele-parallel theory, equivalent of General Relativity, the action is  constructed
by the teleparallel Lagrangian "scalar torsion $T$ ". It   
 has modified versions which are the results of the subtitution of 
scalar torsion in its action  by an arbitrary function of the scalar torsion. This approach is
similar in spirit to the generalization of the Ricci scalar $R$
in the Einstein-Hilbert action to a function $f (R)$.

This theory and its modified versions used orthonormal tetrads defined on the tangent space at each point of the manifold which is the ordinary
space-time. The line element can be written as

\begin{eqnarray}
ds^2=g_{\mu\nu}dx^\mu dx^\nu=\eta_{ij}\theta^i\theta^j\,,
\end{eqnarray}
whose elements can also be expressed  as
\begin{eqnarray}
d^\mu=e_{i}^{\;\;\mu}\theta^{i}; \,\quad \theta^{i}=e^{i}_{\;\;\mu}dx^{\mu}.
\end{eqnarray}
Here $\eta_{ij}=diag(1,-1,-1,-1)$ is the Minkowskian metric and  $\{e^{i}_{\;\mu}\}$ represent the components of tetrads and satisfy the following
identity
\begin{eqnarray}
e^{\;\;\mu}_{i}e^{i}_{\;\;\nu}=\delta^{\mu}_{\nu},\quad e^{\;\;i}_{\mu}e^{\mu}_{\;\;j}=\delta^{i}_{j}.
\end{eqnarray}
We recall here Levi-Civita connection used in General Relativity

\begin{equation}
\overset{\circ }{\Gamma }{}_{\;\;\mu \nu }^{\rho } =
\frac{1}{2}g^{\rho \sigma }\left(
\partial _{\nu} g_{\sigma \mu}+\partial _{\mu}g_{\sigma \nu}-\partial _{\sigma}g_{\mu \nu}\right)\;.
\end{equation}
The curvature associated to this connection is not zero while the torsion vanishes. By opposition to this fundamental property of 
the Levi-Civita connection, the Weizenbock connection  which governs tensor relations in  Tele-Parallel theory and its modified versions
 is defined by
\begin{eqnarray}
\Gamma^{\lambda}_{\mu\nu}=e^{\;\;\lambda}_{i}\partial_{\mu}e^{i}_{\;\;\nu}=-e^{i}_{\;\;\mu}\partial_\nu e_{i}^{\;\;\lambda}.
\end{eqnarray}

With this connection and through the following relations,
we get the  representations of the fundamental  geometrical objects namely the torsion  and contorsion in (\ref{tor}) and (\ref{K2}) respectively,
from which
we determine the tensor $S_{\lambda}^{\;\;\mu\nu}$  in  (\ref{sup})
\begin{eqnarray}\label{tor}
T^{\lambda}_{\;\;\;\mu\nu}= \Gamma^{\lambda}_{\mu\nu}-\Gamma^{\lambda}_{\nu\mu},
\end{eqnarray}

which begots
\begin{equation}\label{K}
K_{\;\;\mu \nu }^{\lambda} \equiv \widetilde{\Gamma} _{\;\mu \nu }^{\lambda }
-\overset{\circ}{\Gamma }{}_{\;\mu \nu }^{\lambda}=\frac{1}{2}(T_{\mu }{}^{\lambda}{}_{\nu }
+ T_{\nu}{}^{\lambda }{}_{\mu }-T_{\;\;\mu \nu }^{\lambda})\;,
\end{equation}

where  $\overset{\circ }{\Gamma }{}_{\;\;\mu \nu }^{\lambda}$ is by definition the Levi-Civita connection. So
\begin{eqnarray}\label{K2}
K^{\mu\nu}_{\;\;\;\;\lambda}=-\frac{1}{2}\left(T^{\mu\nu}_{\;\;\;\lambda}-T^{\nu\mu}_{\;\;\;\;\lambda}+T^{\;\;\;\nu\mu}_{\lambda}\right)\,\,.
\end{eqnarray}
And the tensor $S_{\lambda}^{\;\;\mu\nu}$ is also 

\begin{eqnarray}\label{sup}
S_{\lambda}^{\;\;\mu\nu}=\frac{1}{2}\left(K^{\mu\nu}_{\;\;\;\;\lambda}+
\delta^{\mu}_{\lambda}T^{\alpha\nu}_{\;\;\;\;\alpha}-\delta^{\nu}_{\lambda}T^{\alpha\mu}_{\;\;\;\;\alpha}\right).\label{S}
\end{eqnarray}

The total contraction of torsion tensor by this latter gives the scalar torsion as
\begin{eqnarray}
T=T^{\lambda}_{\;\;\;\mu\nu}S^{\;\;\;\mu\nu}_{\lambda}
\end{eqnarray}
As it has been introduced above, our present investigation will be carried out under the modified theories of 
Tele-Parallel theory where we replace the scalar torsion in Tele-Parallel action   by an arbitrary scalar torsion function, 
giving the action of these modified versions. Indeed, in  the specific case of our present project, the action in a Universe governed by modified 
Tele-Parallel theory is

\begin{eqnarray}
 S= \int e [\frac{T+f(T,\mathcal{T})}{2\kappa^2} +\mathcal{L}_{m} ]d^{4}x,   \label{eq9}
\end{eqnarray}
 where $\kappa^{2} = 8 \pi G $ is the usual gravitational
 coupling constant. By varying the action (\ref{eq9}) with respect
 to the tetrads, one gets the following equations
 of motion \cite{sala1}- \cite{sala4}
\begin{eqnarray}\label{lagran1}
&&[\partial_\xi(ee^\rho_a
S^{\;\;\sigma\xi}_\rho)-ee^\lambda_a S^{\rho\xi\sigma} T_{\rho\xi\lambda}](1+f_T) 
+ e e^\rho_a(\partial_\xi T)S^{\;\;\sigma\xi}_\rho f_{TT} +\frac{1}{4} e e^\sigma_a (T) \nonumber \\
&&  =- \frac{1}{4} e e^\sigma_a \Big( f(\mathcal{T})\Big)  -e e^\rho_a(\partial_\xi \mathcal{T})S^{\;\;\sigma\xi}_\rho f_{T\mathcal{T}}  +
f_{\mathcal{T}}\;\Big(\frac{e\,\Theta^\sigma_{\;\;a}  
+ e e^\sigma_a \;p }{2}\Big) + \frac{\kappa^{2}}{2} e\,\Theta^\sigma_{\;\;a} \;,
\end{eqnarray}
with 
$f_{\mathcal{T}} = \partial f/\partial \mathcal{T} $,   $f_{T} = \partial f/\partial T$, $ f_{T\mathcal{T}} 
= \partial^{2}f/\partial T\partial \mathcal{T}$,
$f_{TT}  = \partial^{2}f/\partial T^{2}$ and  $\Theta^\sigma_{\;\;a}$ the energy-momentum tensor of matter fields.\par 
 After some contraction, we can establish the following relations 
 \begin{eqnarray}\label{nablaS'}
e^a_\nu e^{-1}\partial_\xi(ee^\rho_a
S^{\;\;\sigma\xi}_\rho)-S^{\rho\xi\sigma}T_{\rho\xi\nu} = -\nabla^\xi S_{\nu\xi}^{\;\;\;\;\sigma}-S^{\xi\rho\sigma}K_{\rho\xi\nu}\;,
\end{eqnarray}
and
 \begin{equation}\label{eqdivs'}
G_{\mu\nu}-\frac{1}{2}\,g_{\mu\nu}\,T
=-\nabla^\rho S_{\nu\rho\mu}-S^{\sigma\rho}_{\;\;\;\;\mu}K_{\rho\sigma\nu}\;.
\end{equation}

 We  transform the field equations  Eq.~(\ref{lagran1}) by combining  equations Eq.~(\ref{nablaS'}) and Eq.~(\ref{eqdivs'}). We obtain 
\begin{equation}
A_{\mu\nu}(1+ f_T) +\frac{1}{4}g_{\mu\nu}\;T =B_{\mu\nu}^{eff} \label{motion11},
\end{equation}
with 
\begin{eqnarray}\label{motion1add} 
&&A_{\mu \nu }=g_{\sigma\mu}e^a_\nu[e^{-1}\partial_\xi(ee^\rho_a
S^{\;\;\sigma\xi}_\rho)-e^\lambda_a S^{\rho\xi\sigma} T_{\rho\xi\lambda}],\\\nonumber
&&\qquad=-\nabla^\sigma S_{\nu\sigma\mu }-S_{\;\;\;\;\mu }^{\rho\lambda }K_{\lambda \rho \nu }
=G_{\mu \nu }-\frac{1}{2}g_{\mu \nu }T, \; \\\nonumber
&&B_{\mu\nu}^{eff} =  S^{\rho}_{\;\;\;\mu\nu}\; f_{T\mathcal{T}}\; \partial_{\rho} \mathcal{T}  -
S^{\rho}_{\;\;\;\mu\nu}\;f_{TT}\; \partial_{\rho} T  
  - \frac{1}{4} g_{\mu \nu }f +
f_{\mathcal{T}}\;\Big(\frac{\Theta_{\mu \nu }  
+ g_{\mu \nu } \;p }{2}\Big) + \frac{\kappa^{2}}{2} \,\Theta_{\mu \nu } \;.
\end{eqnarray}
Consequently, we can rewrite the equation Eq.~(\ref{motion11}) in  the following form
\begin{equation}
(1+ f_T)\,G_{\mu\nu}=T_{\mu\nu}^{eff} \label{motion12},
\end{equation}
where
\begin{eqnarray}\label{motion1add'} 
T_{\mu\nu}^{eff} =S^{\rho}_{\;\;\;\mu\nu}\; f_{T\mathcal{T}}\; \partial_{\rho} \mathcal{T}  -
S^{\rho}_{\;\;\;\mu\nu}\;f_{TT}\; \partial_{\rho} T  - \frac{1}{4} g_{\mu \nu } \Big(T+ f \Big) +\frac{T\,g_{\mu\nu}\,f_T}{2}+
f_{\mathcal{T}}\;\Big(\frac{\Theta_{\mu \nu }  
+ g_{\mu \nu } \;p }{2}\Big) + \frac{\kappa^{2}}{2} \,\Theta_{\mu \nu } \;.
\end{eqnarray}

 \section{Geodesic Deviation Equation in GR}\label{sec3}
In this section, we revisited briefly the GDE notions in General Relativity. Indeed, in order  to explain well
the geometrical meaning of Riemann tensor,
it would be necessary to look into the behavior of two neighboring geodesics. So, let's assume two neighboring geodesics $C_{1}$ and  $C_{2}$
with affine parameter $\nu$ on a 2-surface  $S$ (see fig.1). 
The field vector $V^{\alpha}=\frac{dx^{\alpha}}{d\nu}$ is the  normalized  tangent vector of the geodesic 
 $C_{1}$ and  $\eta^{\alpha}=\frac{dx^{\alpha}}{ds}$  is the deviation vector of these two geodesics. Therefore, we characterize these geodesics  
 with $x^{\alpha}(\nu,s)$.\\
 Beginning with $\pounds_{_{V}}\eta^{\alpha}=\pounds_{_{\eta}}V^{\alpha}$ ($[V,\eta]^{\alpha}=0$) which leads to  
 $\nabla_{_{V}}\nabla_{_{V}}\eta^{\alpha}=\nabla_{_{V}}\nabla_{\eta}V^{\alpha}$ and using
 $\nabla_{_{X}}\nabla_{_{Y}}Z^{\alpha}-\nabla_{_{Y}}\nabla_{X}Z^{\alpha}-\nabla_{_{[X,Y]}}Z^{\alpha}=
R^{\alpha}\,_{\beta\gamma\delta}Z^{\beta}X^{\gamma}Y^{\delta}$ in which  $Y^{\alpha}=\eta^{\alpha}$ and $X^{\alpha}=Z^{\alpha}=V^{\alpha}$, we can 
obtain the GDE as doing in \cite{3} by 

\begin{equation}\label{salako12}
\frac{D^{2}\eta^{\alpha}}{D\nu^{2}}=-R^{\alpha}\,_{\beta\gamma\delta}V^{\beta}\eta^{\gamma}V^{\delta}.
\end{equation}
Now,  we briefly review the research results on the GDE in GR. We use the energy-momentum tensor in a purportedly  perfect fluid 

\begin{equation}\label{salako14}
\Theta_{\mu\nu}=(\rho+p)u_{\alpha}u_{\beta}+pg_{\alpha\beta},
\end{equation}
 where $\rho$ and  $p$  denote respectivily the energy density and the pressure. We recall here that the trace of energy-momentum tensor is given by
\begin{equation}\label{salako15}
\mathcal{T} = \rho - 3p.
\end{equation}
Considering  the Einstein field equations in GR (with cosmological constant) given by
\begin{equation}\label{salako16}
R_{\mu\nu}-\frac{1}{2}Rg_{\mu\nu}+\Lambda g_{\mu\nu}=\kappa\Theta_{\mu\nu},
\end{equation}
   we can determinate   the  Ricci scalar and Ricci tensor  as follows
\begin{equation}\label{salako17}
R=\kappa (\rho-3p)+4\Lambda,
\end{equation}

\begin{equation}\label{salako18}
R_{\mu\nu}=\kappa (\rho+p)u_{\alpha}u_{\beta}+\frac{1}{2}[\kappa(\rho-p)+2\Lambda]g_{\mu\nu}.
\end{equation}
And then, we determinate the   (0-4) Riemann curvature tensor by using Eq.(\ref{salako17}) and  Eq.(\ref{salako18}). It results 
\begin{eqnarray}\label{salako19}
R_{\alpha\beta\gamma\delta}&=&\frac{1}{2}(g_{\alpha\gamma}R_{\delta\beta}-
g_{\alpha\delta}R_{\gamma\beta}+g_{\beta\delta}R_{\gamma\alpha}-g_{\beta\gamma}R_{\delta\alpha})- \frac{R}{6}(g_{\alpha\gamma}g_{\delta\beta}-
g_{\alpha\delta}g_{\gamma\beta})+C_{\alpha\beta\gamma\delta}.
\end{eqnarray}
Thus, the right hand side of Eq.(\ref{salako12}) becomes
\begin{eqnarray}\label{salako20}
R^{\alpha}\,_{\beta\gamma\delta}V^{\beta}\eta^{\gamma}V^{\delta}=\left[\frac{1}{3}(\kappa \rho+\Lambda)\epsilon +\frac{1}{2}\kappa (\rho+p)E^{2}\right]\eta^{\alpha},
\end{eqnarray}
where  $\epsilon=V^{\alpha}V_{\alpha}$ and $E=-V^{\alpha}u_{\alpha}$. We emphasize here that   $\epsilon=(-1, 0 ,1)$, which corresponds 
to  space-like, null and 
time-like  geodesics respectively . The equation Eq.(\ref{salako20}) reminds us {\it  Pirani equation } \cite{2}. We note  that by considering null the Weyl tensor, 
we can extract the Pirani equation for every metric, which  makes this equation, an equation giving solutions for space-like, null and timelike 
congruences 
\cite{ellis}.  In  this work, we generalize these resultats in the framework of  $f(T,\mathcal{T})$ gravity applied to the metric.  
\begin{figure}[ht]
  \centering
  \includegraphics[width=4cm]{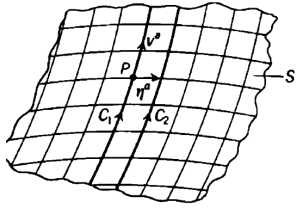}~
\caption{Geodesic Deviation..}
  \label{fig:WEC1a}
\end{figure}
\section{Geodesics Deviation Equation in    $f(T,\mathcal{T})$ gravity}\label{sec4}
In this section, we establish the GDE in the context of modified $f(T,\mathcal{T})$ gravity. Indeed, we begin our investigation 
by the Ricci scalar determination from the trace of equation  Eq.(\ref{motion11}),  which leads to    
\begin{eqnarray}\label{salako21}
R= \frac{-1}{(1+f_T)} \Bigg[ S^{\mu}_{\;\;\;\mu\rho}\; f_{T\mathcal{T}}\; \partial^{\rho} \mathcal{T}  -
 S^{\mu}_{\;\;\;\mu\rho}\;f_{TT}\; \partial^{\rho} T  +   \Big(T- f(T,\mathcal{T}) \Big) +
f_{\mathcal{T}}\;\Big(\frac{\Theta  
+4\;p }{2}\Big) + \frac{\kappa^{2}}{2} \,\Theta\Bigg].
\end{eqnarray} 
In  order to put out the Ricci tensor in the framework of the  $f(T,\mathcal{T})$ gravity, one follows the same way as it has been done in GR 
by introducing Eq.(\ref{salako21}) in Eq.(\ref{motion11}). Thus the modified Ricci tensor is presented as follow
\begin{eqnarray}\label{salako22}
R_{\mu\nu}&=&\frac{1}{(1+f_{T})}\Bigg\{ - g_{\mu\nu} \Big[ S^{\mu}_{\;\;\;\mu\rho}\; f_{T\mathcal{T}}\; \partial^{\rho} \mathcal{T}  -
 S^{\mu}_{\;\;\;\mu\rho}\;f_{TT}\; \partial^{\rho} T  +   \Big(T- f(T,\mathcal{T}) \Big) +
f_{\mathcal{T}}\;\Big(\frac{\Theta  
+4\;p }{2}\Big) \cr
&&+ \frac{\kappa^{2}}{2} \,\Theta -\frac{1}{4}  (T- f(T,\mathcal{T}) )-\frac{f_{\mathcal{T}}\;p }{2}\Big]\cr
&&+ S^{\rho}_{\;\;\;\mu\nu}\; f_{T\mathcal{T}}\; \partial_{\rho} \mathcal{T}  -
S^{\rho}_{\;\;\;\mu\nu}\;f_{TT}\; \partial_{\rho} T   + \frac{(\kappa^{2}+f_{\mathcal{T}})}{2} \,\Theta_{\mu \nu }  \Bigg\}.
\end{eqnarray}
If we suppose vanish the Weyl tensor $C_{\alpha\beta\gamma\delta}$,  Eq.(\ref{salako19})  becomes 
\begin{eqnarray}\label{salako23}
R_{\alpha\beta\gamma\delta}&=&\frac{1}{2(1+f_{T})}\Big[\frac{(\kappa^{2}+f_{\mathcal{T}})}{2}(g_{\alpha\gamma}
\Theta_{\delta\beta}-g_{\alpha\delta}\Theta_{\gamma\beta}+
g_{\beta\delta}\Theta_{\gamma\alpha}-g_{\beta\gamma}\Theta_{\delta\alpha})\cr
&&
- g_{\mu\nu} \Big[ S^{\mu}_{\;\;\;\mu\rho}\; f_{T\mathcal{T}}\; \partial^{\rho} \mathcal{T}  -
 S^{\mu}_{\;\;\;\mu\rho}\;f_{TT}\; \partial^{\rho} T  +   \Big(T- f(T,\mathcal{T}) \Big) +
f_{\mathcal{T}}\;\Big(\frac{\Theta  
+4\;p }{2}\Big) \cr
&&+ \frac{\kappa^{2}}{2} \,\Theta -\frac{1}{4}  (T- f(T,\mathcal{T}) )-\frac{f_{\mathcal{T}}\;p }{2}\Big]\cr
&& 
\times   (g_{\alpha\gamma} g_{\delta\beta}-g_{\alpha\delta}g_{\gamma\beta}) +
(g_{\alpha\gamma}D_{\delta\beta}-
g_{\alpha\delta}D_{\gamma\beta}+g_{\beta\delta}D_{\gamma\alpha}-g_{\beta\gamma}D_{\delta\alpha})f_{T}\cr
&&
+
(g_{\alpha\gamma}\mathcal{D}_{\delta\beta}-
g_{\alpha\delta}\mathcal{D}_{\gamma\beta}+g_{\beta\delta}\mathcal{D}_{\gamma\alpha}-g_{\beta\gamma}\mathcal{D}_{\delta\alpha})f_{T }
\Big]\cr
&&-
\frac{1}{6(1+f_{T})}\Bigg\{ - g_{\mu\nu} \Big[ S^{\mu}_{\;\;\;\mu\rho}\; f_{T\mathcal{T}}\; \partial^{\rho} \mathcal{T}  -
 S^{\mu}_{\;\;\;\mu\rho}\;f_{TT}\; \partial^{\rho} T  +   \Big(T- f(T,\mathcal{T}) \Big) +
f_{\mathcal{T}}\;\Big(\frac{\Theta  
+4\;p }{2}\Big) \cr
&&+ \frac{\kappa^{2}}{2} \,\Theta -\frac{1}{4}  (T- f(T,\mathcal{T}) )-\frac{f_{\mathcal{T}}\;p }{2}\Big]+ S^{\rho}_{\;\;\;\mu\nu}\; 
f_{T\mathcal{T}}\; \partial_{\rho} \mathcal{T}  -
S^{\rho}_{\;\;\;\mu\nu}\;f_{TT}\; \partial_{\rho} T  \cr
&&+ \frac{(\kappa^{2}+f_{\mathcal{T}})}{2} \,\Theta_{\mu \nu }  \Bigg\}
(g_{\alpha\gamma}
g_{\delta\beta}-g_{\alpha\delta}g_{\gamma\beta}),
\end{eqnarray}
where
\begin{equation}\label{salako24}
D_{\mu\nu}=-S_{\nu\mu\rho} \nabla^{\rho}T \partial_{_{T}},
\end{equation}
and
\begin{equation}\label{salako24'}
\mathcal{D}_{\mu\nu}=-S_{\nu\mu\rho} \nabla^{\rho}\,\mathcal{T} \partial_{_{\mathcal{T}}}.
\end{equation}
Now, we rise the first  index $ \alpha $  of Riemann tensor in Eq.(\ref{salako19}) by a contraction with $V^{\beta}\eta^{\gamma}V^{\delta}$, leading 
to the following  (1, 3)-tensor 
\begin{eqnarray}\label{salako25}
R^{\alpha}\,_{\beta\gamma\delta}V^{\beta}\eta^{\gamma}V^{\delta}&=& \frac{1}{2(1+f_{T})}\Big[\frac{(\kappa^{2}+f_{\mathcal{T}})}{2}
(\delta^{\alpha}_{\gamma}\Theta_{\delta\beta}-
\delta^{\alpha}_{\delta}\Theta_{\gamma\beta}
+g_{\beta\delta}\Theta^{\alpha}_{\gamma}-g_{\beta\gamma}\Theta^{\alpha}_{\delta})\\\nonumber
&&
-\Big[ S^{\mu}_{\;\;\;\mu\rho}\; f_{T\mathcal{T}}\; \partial^{\rho} \mathcal{T}  -
 S^{\mu}_{\;\;\;\mu\rho}\;f_{TT}\; \partial^{\rho} T  +   \Big(T- f(T,\mathcal{T}) \Big) +
f_{\mathcal{T}}\;\Big(\frac{\Theta  
+4\;p }{2}\Big) \cr
&&+ \frac{\kappa^{2}}{2} \,\Theta -\frac{1}{4}  (T- f(T,\mathcal{T}) )-\frac{f_{\mathcal{T}}\;p }{2}\Big]\cr
&& 
\times   (\delta^{\alpha}_{\gamma}g_{\delta\beta}-\delta^{\alpha}_{\delta}g_{\gamma\beta}) +
(\delta^{\alpha}_{\gamma}D_{\delta\beta}-\delta^{\alpha}_{\delta}D_{\gamma\beta}+
g_{\beta\delta}D^{\alpha}_{\gamma}-g_{\beta\gamma}D^{\alpha}_{\delta})f_{T}    \cr
&&
+(\delta^{\alpha}_{\gamma}\mathcal{D}_{\delta\beta}-\delta^{\alpha}_{\delta}\mathcal{D}_{\gamma\beta}+
g_{\beta\delta}\mathcal{D}^{\alpha}_{\gamma}-g_{\beta\gamma}\mathcal{D}^{\alpha}_{\delta})f_{T}   \Big]V^{\beta}\eta^{\gamma}V^{\delta}
\cr
&&-
\frac{1}{6(1+f_{T})}\Bigg\{ - g_{\mu\nu} \Big[ S^{\mu}_{\;\;\;\mu\rho}\; f_{T\mathcal{T}}\; \partial^{\rho} \mathcal{T}  -
 S^{\mu}_{\;\;\;\mu\rho}\;f_{TT}\; \partial^{\rho} T  +   \Big(T- f(T,\mathcal{T}) \Big) +
f_{\mathcal{T}}\;\Big(\frac{\Theta  
+4\;p }{2}\Big) \cr
&&+ \frac{\kappa^{2}}{2} \,\Theta -\frac{1}{4}  (T- f(T,\mathcal{T}) )-\frac{f_{\mathcal{T}}\;p }{2}\Big]+ S^{\rho}_{\;\;\;\mu\nu}\; 
f_{T\mathcal{T}}\; \partial_{\rho} \mathcal{T}  -
S^{\rho}_{\;\;\;\mu\nu}\;f_{TT}\; \partial_{\rho} T   + \frac{(\kappa^{2}+f_{\mathcal{T}})}{2} \,\Theta_{\mu \nu }  \Bigg\}\cr
&&
(g_{\alpha\gamma}g_{\delta\beta}-
g_{\alpha\delta}g_{\gamma\beta})V^{\beta}\eta^{\gamma}V^{\delta}.
\end{eqnarray}
We remark here that Eq.(\ref{salako23}) and Eq.(\ref{salako25}) are available only if the Weyl tensor is reduced to zero. 
Otherwise, from  Eqs(\ref{salako14}) and (\ref{salako23}), one points out the following relation

\begin{eqnarray}\label{salako26}
R^{\alpha}\,_{\beta\gamma\delta}&=&\frac{1}{2(1+f_{T})}
\Bigg\{\frac{(\kappa^{2}+f_{\mathcal{T}})}{2}(\rho+p)(g_{\alpha\gamma}u_{\delta}u_{\beta}-g_{\alpha\delta}u_{\gamma}u_{\beta}+
g_{\beta\delta}u_{\gamma}u_{\alpha}-g_{\beta\gamma}u_{\delta}u_{\alpha})-
(g_{\alpha\gamma}g_{\delta\beta}-g_{\alpha\delta}g_{\gamma\beta})\cr
&& \times \frac{1}{3}
\Big[ S^{\mu}_{\;\;\;\mu\rho}\; f_{T\mathcal{T}}\; \partial^{\rho} \mathcal{T}  -
 S^{\mu}_{\;\;\;\mu\rho}\;f_{TT}\; \partial^{\rho} T  +   \Big(T- f(T,\mathcal{T}) \Big) +
f_{\mathcal{T}}\;\Big(\frac{\Theta  
+4\;p }{2}\Big) + \frac{\kappa^{2}}{2} \,\Theta -\frac{1}{4}  (T- f(T,\mathcal{T}) )\cr
&&
-\frac{f_{\mathcal{T}}\;p }{2}\Big]\cr
&&
+(g_{\alpha\gamma}D_{\delta\beta}-g_{\alpha\delta}D_{\gamma\beta}+
g_{\beta\delta}D_{\gamma\alpha}-g_{\beta\gamma}D_{\delta\alpha})f_{T}+ (g_{\alpha\gamma}\mathcal{D}_
{\delta\beta}-g_{\alpha\delta}\mathcal{D}_{\gamma\beta}+
g_{\beta\delta}\mathcal{D}_{\gamma\alpha}-g_{\beta\gamma}\mathcal{D}_{\delta\alpha})f_{T}\Bigg\}.
\end{eqnarray}
Under the condition of vector field normalization, we have $V^{\alpha}V_{\alpha}=\epsilon$ and
\begin{eqnarray}\label{salako27}
R_{\alpha\beta\gamma\delta}V^{\beta}V^{\delta}&=&\frac{1}{2(1+f_{T})}\Bigg\{\frac{(\kappa^{2}+f_{\mathcal{T}})}{2}(\rho+p)(g_{\alpha\gamma}
(u_{\beta}V^{\beta})^{2}-2(u_{\beta}V^{\beta})V_{(\alpha}u_{\gamma)}+\epsilon u_{\alpha}u_{\gamma})\cr
&&-
 \frac{1}{3}
\Big[ S^{\mu}_{\;\;\;\mu\rho}\; f_{T\mathcal{T}}\; \partial^{\rho} \mathcal{T}  -
 S^{\mu}_{\;\;\;\mu\rho}\;f_{TT}\; \partial^{\rho} T  +   \Big(T- f(T,\mathcal{T}) \Big) +
f_{\mathcal{T}}\;\Big(\frac{\Theta  
+4\;p }{2}\Big) \cr
&&+ \frac{\kappa^{2}}{2} \,\Theta -\frac{1}{4}  (T- f(T,\mathcal{T}) )-\frac{f_{\mathcal{T}}\;p }{2}\Big]\cr
&& \times
(\epsilon g_{\alpha\gamma}-V_{\alpha}V_{\gamma})  +
[(g_{\alpha\gamma}D_{\delta\beta}-g_{\alpha\delta}D_{\gamma\beta}
+g_{\beta\delta}D_{\gamma\alpha}-g_{\beta\gamma}D_{\delta\alpha})f_{T}]V^{\beta}V^{\delta}\cr
&&
+ 
[(g_{\alpha\gamma}\mathcal{D}_{\delta\beta}-g_{\alpha\delta}\mathcal{D}_{\gamma\beta}
+g_{\beta\delta}\mathcal{D}_{\gamma\alpha}-g_{\beta\gamma}\mathcal{D}_{\delta\alpha})f_{T}]V^{\beta}V^{\delta}
\Bigg\}.
\end{eqnarray}
 We rise the first index with  $\eta^{\gamma}$ and obtain
\begin{eqnarray}\label{salako28}
R^{\alpha}\,_{\beta\gamma\delta}V^{\beta}\eta^{\gamma}V^{\delta}&=&\frac{1}{2(1+f_{T})}\Big\{\frac{(\kappa^{2}+f_{\mathcal{T}})}{2}(\rho+p)
((u_{\beta}V^{\beta})^{2}\eta^{\alpha}-
(u_{\beta}V^{\beta})V^{\alpha}(u_{\gamma}\eta^{\gamma})\cr
&&-(u_{\beta}V^{\beta})u^{\alpha}(V_{\gamma}\eta^{\gamma})
+\epsilon u_{\alpha}u_{\gamma}\eta^{\gamma})\cr
&&-
\frac{1}{3}
\Big[ S^{\mu}_{\;\;\;\mu\rho}\; f_{T\mathcal{T}}\; \partial^{\rho} \mathcal{T}  -
 S^{\mu}_{\;\;\;\mu\rho}\;f_{TT}\; \partial^{\rho} T  +   \Big(T- f(T,\mathcal{T}) \Big) +
f_{\mathcal{T}}\;\Big(\frac{\Theta  
+4\;p }{2}\Big)\cr
&&+ \frac{\kappa^{2}}{2} \,\Theta -\frac{1}{4}  (T- f(T,\mathcal{T}) )-\frac{f_{\mathcal{T}}\;p }{2}\Big]\cr
&& \times
(\epsilon \eta^{\alpha}-V_{\alpha}(V_{\gamma}\eta^{\gamma}))
+[(\delta^{\alpha}_{\gamma}D_{\delta\beta}-\delta^{\alpha}_{\delta}D_{\gamma\beta}+g_{\beta\delta}D^{\alpha}_{\gamma}-
g_{\beta\gamma}D^{\alpha}_{\delta})f_{T}]V^{\beta}V^{\delta}\eta^{\gamma}  
\cr
&&
+[(\delta^{\alpha}_{\gamma}\mathcal{D}_{\delta\beta}-\delta^{\alpha}_{\delta}\mathcal{D}_{\gamma\beta}
+g_{\beta\delta}\mathcal{D}^{\alpha}_{\gamma}-
g_{\beta\gamma}\mathcal{D}^{\alpha}_{\delta})f_{T}]V^{\beta}V^{\delta}\eta^{\gamma} 
\Big\}.
\end{eqnarray}
By  using $E=-V_{\alpha}u^{\alpha}$ and $\eta_{\alpha}u^{\alpha}=\eta_{\alpha}V^{\alpha}=0$,
Eq. (\ref{salako28}) becomes
\begin{eqnarray}\label{salako29}
R^{\alpha}\,_{\beta\gamma\delta}V^{\beta}\eta^{\gamma}V^{\delta}&=&\frac{1}{2(1+f_{T})}
\Big[\frac{(\kappa^{2}+f_{\mathcal{T}})}{2}(\rho+p)E^{2}-
\frac{\epsilon}{3}
\Big( S^{\mu}_{\;\;\;\mu\rho}\; f_{T\mathcal{T}}\; \partial^{\rho} \mathcal{T}  -
 S^{\mu}_{\;\;\;\mu\rho}\;f_{TT}\; \partial^{\rho} T  +   \Big(T- f(T,\mathcal{T}) \Big) \cr
 &&+
f_{\mathcal{T}}\;\Big(\frac{\Theta  
+4\;p }{2}\Big) + \frac{\kappa^{2}}{2} \,\Theta -\frac{1}{4}  (T- f(T,\mathcal{T}) )-\frac{f_{\mathcal{T}}\;p }{2}\Big)
\Big]\eta^{\alpha}\cr
&&+
\frac{1}{2(1+f_{T})}\Big[(\delta^{\alpha}_{\gamma}D_{\delta\beta}-\delta^{\alpha}_{\delta}D_{\gamma\beta}+
g_{\beta\delta}D^{\alpha}_{\gamma}-
g_{\beta\gamma}D^{\alpha}_{\delta})f_{T}]V^{\beta}V^{\delta}\Big]\eta^{\gamma}\cr
&&+
\frac{1}{2(1+f_{T})}\Big[(\delta^{\alpha}_{\gamma}\mathcal{D}_{\delta\beta}-\delta^{\alpha}_{\delta}
\mathcal{D}_{\gamma\beta}+g_{\beta\delta}\mathcal{D}^{\alpha}_{\gamma}-
g_{\beta\gamma}\mathcal{D}^{\alpha}_{\delta})f_{T}]V^{\beta}V^{\delta}\Big]\eta^{\gamma}.
\end{eqnarray}
In the following section and by make using  the FLRW metric whose Weyl tensor is zero;
we will use these previous results to obtain the  GDE in the framework of  $ f (T, \mathcal {T}) $ gravity, which is of course the
equivalent of GDE in GR.
\subsection{  GR equivalent method with FLRW background}
Here we are interested in studying flat FLRW cosmologies whose metric can be described by, 
\begin{equation}\label{salako13}
 ds^{2}=-dt^{2}+a^{2}(t)\left[\frac{dr^{2}}{1-k r^{2}}+r^{2}d\theta^{2}+r^{2}sin^{2}\theta d\phi^{2}\right].
\end{equation}
The Weyl tensor associated to this latter is zero because of the flatness conformal of Universe described by FLRW metric. 
  The non-zero components of tetrads according to the above metric  are given by  
\begin{eqnarray}
 \{e^{a}_{\;\; \mu}\}= diag[1,a,a,a]. \label{eq11}
\end{eqnarray}
The determinant of the matrix (\ref{eq11}) is $e = a^{3}$ and the  non-zero components of the torsion tensor and contorsion
 tensor are also given by
\begin{eqnarray}
 T^{1}_{\;\;\; 01}= T^{2}_{\;\;\; 02}=T^{3}_{\;\;\; 03}=\frac{\dot{a}}{a},\\
 K^{01}_{\;\;\;\;1}=K^{ 02}_{\;\;\;\;2}=K^{ 03}_{\;\;\;\;3}= \frac{\dot{a}}{a}. \label{eq12}
\end{eqnarray}
 The  non-zero components of the  tensor $S^{\;\;\; \mu\nu}_{\alpha}$  are 
\begin{eqnarray}
S^{\;\;\; 11}_{0}=S^{\;\;\; 22}_{0}=S^{\;\;\; 33}_{0}=\frac{\dot{a}}{a}.  \label{eq12'}
\end{eqnarray}
 Therefore, one evaluates  the torsion scalar and obtains 
\begin{equation}\label{salako30}
T=-6H^{2},
\end{equation}
where $H=\dot{a}/a$ denotes the Hubble parameter. The normalization of the vector field leads to  $V_{\alpha}V^{\alpha}=\epsilon$ and we have also  
$E=-V_{\alpha}u^{\alpha}$, $\eta_{\alpha}u^{\alpha}=\eta_{\alpha}V^{\alpha}=0$, $\eta_{0}u^{0}=0$.
Considering (\ref{eq12}) and (\ref{eq12'}) we find $S^{1}_{10}=S^{2}_{20}=S^{3}_{30}=-2H(t)$. 
Thus, we can reduce the  expression of  $R^{\alpha}\,_{\beta\gamma\delta}V^{\beta}\eta^{\gamma}V^{\delta}$ as follows

\begin{eqnarray}\label{salako32}
R^{\alpha}\,_{\beta\gamma\delta}V^{\beta}\eta^{\gamma}V^{\delta}&=& \frac{G}{(1+f_T)}\left(1+\frac{f_{\mathcal{T}}}{16\pi G}\right)
\Bigg\{ 4\pi E^2 (\rho +\rho_1+p+p_1)\cr
&&+ \frac{8\pi\,\epsilon}{3} (\rho +\rho_1)   \Bigg \}\eta^{\alpha}
\end{eqnarray}
with  
\begin{eqnarray}
&&G_{eff}=\frac{G}{(1+f_T)}\left(1+\frac{f_{\mathcal{T}}}{16\pi G}\right)\label{Geff}\,,\\
&&\rho_1=\frac{1}{16\pi G_{eff}(1+f_T)}\left(f_{\mathcal{T}}p-\frac{1}{2}f- \frac{T}{2} \right)\label{densDE}\,,\\
&&p_1=\rho+p-\rho_1-\frac{1}{4\pi G_{eff}(1+f_T)}\left[12H^2\dot{H}f_{TT}-f_{T\mathcal{T}}H\left(\dot{\rho}-
3\dot{p}\right)\right]\label{pDE}\,.
\end{eqnarray}
The relation so posed  in (\ref{salako32}) is  the generalized {\it Pirani} equation.
Now, we can write the GDE in $f(T,\mathcal{T})$ gravity model as following
\begin{eqnarray}\label{salako33}
\frac{D^{2}\eta^{\alpha}}{D\nu^{2}}&=&- \frac{G}{(1+f_T)}\left(1+\frac{f_{\mathcal{T}}}{16\pi G}\right)
\Bigg\{ 4\pi E^2 (\rho +\rho_1+p+p_1)     + \frac{8\pi\,\epsilon}{3} (\rho +\rho_1)   \Bigg \}\eta^{\alpha}.
\end{eqnarray}
For reasons of homogeneity and isotropy of the metric FLRW, there has been a change in the intensity of the deviation vector $\eta^{\alpha}$ where as 
 in the anisotropic Universe like Bianchi Universe $\textsc{I}$, the GDE has also induced a change in the
direction of the deviation vector as described in  \cite{cac}.

 \subsection{Direct method with FLRW background }
 In this section,  we can write the GDE in $f(T,\mathcal{T})$ gravity model by evaluating the  LHS of Eq.(\ref{salako32}) 
  for $\alpha=r$ as follows
\begin{equation}\label{salako411}
R^{r}\,_{\beta\gamma\delta}V^{\beta}\eta^{\gamma}V^{\delta}=R^{r}\,_{t\gamma t}V^{t}\eta^{\gamma}V^{t}+R^{r}
\,_{r\gamma r}V^{r}\eta^{\gamma}V^{r}+
R^{r}\,_{\theta \gamma  \theta}V^{\theta}\eta^{\gamma}V^{\theta}+R^{r}\,_{\phi \gamma  \phi}V^{\phi}\eta^{\gamma}V^{\phi}.
\end{equation}
Assuming that the Riemann tensor components are not different to zero for  FLRW metric in Eq.(\ref{salako13}), we can take $\gamma=r$.
Indeed, Eq.(\ref{salako411}) becames:
\begin{equation}\label{salako412}
R^{r}\,_{\beta\gamma\delta}V^{\beta}\eta^{\gamma}V^{\delta}=R^{r}\,_{trt}V^{t}V^{t}\eta^{r}+R^{r}\,_{rrr}~g^{rr}V_{r}V^{r}\eta^{r}+
R^{r}\,_{\theta r\theta}~g^{\theta\theta}V_{\theta}V^{\theta}\eta^{r}+
R^{r}\,_{\phi r\phi}~g^{\phi\phi}V_{\phi}V^{\phi}\eta^{r}=(-\dot{H}E^{2}+\epsilon H^{2}) \eta^{r},
\end{equation}
 where we have used the following expressions
$V^{t}V^{t}=E^{2}$, $V_{i}\eta^{i}=0$, $R^{r}\,_{rrr}=0$, $R^{r}\,_{\theta r\theta}=
r^{2}\dot{a}^{2}$, $R^{r}\,_{\phi r\phi}=\dot{a}^{2}r^{2}sin^{2}\theta$ and $R^{r}\,_{ttr}=\frac{\ddot{a}}{a}$. 
Remark that similar equations are also obtained for $\alpha=\theta$ and $\alpha=\phi$.
Considering (\ref{lagran1}) and (\ref{salako13}), we obtain the  Friedmann standard equations
\begin{equation}\label{salakoee}
3H^2=8\pi G_{eff}\left(\rho+\rho_1\right),
\end{equation}
 and
\begin{equation}\label{salakoeee}
\dot{H}=-4\pi G_{eff}\left(\rho+p+\rho_1+p_1\right).
\end{equation}
By putting the equations (\ref{salakoee}) and  (\ref{salakoeee}) in equation (\ref{salako412}), we can generalize the {\it Pirani} equations by 
a direct approach as 
\begin{eqnarray}\label{salakoeeee}
R^{\alpha}\,_{\beta\gamma\delta}V^{\beta}\eta^{\gamma}V^{\delta}&=& \frac{G}{(1+f_T)}\left(1+\frac{f_{\mathcal{T}}}{16\pi G}\right)
\Bigg\{ 4\pi E^2 (\rho +\rho_1+p+p_1)\cr
&&+ \frac{8\pi\,\epsilon}{3} (\rho +\rho_1)   \Bigg \}\eta^{\alpha},
\end{eqnarray}
 which leads to the same Geodesic Deviation Equation as in (\ref{salako33}). This means that we have found
the same results by two different approaches . This result proves the validity of the GDE obtained in the context of $f(T,\mathcal{T})$ gravity.
  \subsection{ Fundamental observers with FLRW background}
Here, we are basing  our  analysis on the fundamental observers. In this particular case, we interprete   $V^{\alpha}$ and  $\nu$ 
(affine parameter)  as the four-viscosity of fluid $u^{\alpha}$ and  $t$ ( proper time). Since we are performing  with temporal geodesics, we have
$\epsilon=1$. By constraining  the vector field normalization as  $ E = 1$, one gets  

\begin{eqnarray}\label{salako34}
R^{\alpha}\,_{\beta\gamma\delta}u^{\beta}\eta^{\gamma}u^{\delta}=
\frac{4\pi\,G}{(1+f_T)}\left(1+\frac{f_{\mathcal{T}}}{16\pi G}\right)
\Bigg\{\frac{\rho+\rho_1}{3} +(p+p_1)  \Bigg \}\eta^{\alpha}.
\end{eqnarray}
 We know that if $\eta_{\alpha}=\ell e_{\alpha}$, where $e_{\alpha}$ is parallel propagated along $t$, then the isotropy results in
\begin{eqnarray}\label{salako35}
\frac{De^{\alpha}}{Dt}=0,
\end{eqnarray}
 which induces 
\begin{eqnarray}\label{salako36}
\frac{D^{2}\eta^{\alpha}}{Dt^{2}}=\frac{d^{2}\ell}{dt^{2}}e^{\alpha}.
\end{eqnarray}
 By using (\ref{salako12}) and (\ref{salako34}) we can write
\begin{eqnarray}\label{salako37}
\frac{d^{2}\ell}{dt^{2}}=\frac{-4\pi\,G}{(1+f_T)}\left(1+\frac{f_{\mathcal{T}}}{16\pi G}\right)
\Bigg\{\frac{\rho+\rho_1}{3} +(p+p_1)  \Bigg \} \ell.
\end{eqnarray}
For the particular case $\ell=a(t)$, equation (\ref{salako37}) becames
\begin{eqnarray}\label{salako38}
\frac{\ddot{a}}{a}=    \frac{-4\pi\,G}{(1+f_T)}\left(1+\frac{f_{\mathcal{T}}}{16\pi G}\right)
\Bigg\{\frac{\rho+\rho_1}{3} +(p+p_1)  \Bigg \}.
\end{eqnarray}
 This equation is a particular case of the generalized Raychaudhuri equation 
 given in \cite{eti20} .
Furthermore, from the standard forms of the modified  Friedmann equations in $f(T,\mathcal{T})$ gravity model for flat Universe \cite{rafael2}
, the  generalized Raychaudhuri equation above can be obtained. These equations are expressed as follows
\begin{eqnarray}
H^{2}&=&\frac{8\pi G}{3} \rho-\frac{1}{6}\left(f+12 H^{2} f_{T}\right)+f_{%
\mathcal{T}} \left(\frac{\rho+p}{3}\right),\label{salako39}\cr
\dot{H}&=&-\frac{4\pi G\left( 1+f_{\mathcal{T}}/8\pi G\right) \left( \rho+p\right) }{1+f_{T}-12H^{2}f_{TT}+H\left( d\rho /dH\right)
\left( 1-3c_{s}^{2}\right)f_{T\mathcal{T}} }\label{salako40}.
\end{eqnarray} 
 Consistency between the modified Friedmann equations in  $ f (T, \mathcal {T}) $  gravity applied to  flat Universe \cite{rafael2} and  the
  generalized  Raychaudhuri equation for  flat Universe (\ref{salako38}) confirms that the approach  followed  here, 
  is one of the valid approaches.
 
  \subsection{ Null vector fields with FLRW background}

 In this section,  we suppose that vector fields are  directed the null past,  namely  $V^{\alpha}=k^{\alpha}, k_{\alpha}k^{\alpha}=0$,
 for which the Eq.(\ref{salako32}) leads to

\begin{eqnarray}\label{salako41}
R^{\alpha}\,_{\beta\gamma\delta}k^{\beta}\eta^{\gamma}k^{\delta}=
\frac{4\pi\,G}{(1+f_T)}\left(1+\frac{f_{\mathcal{T}}}{16\pi G}\right)
\Bigg\{ (\rho +\rho_1+p+p_1)  \Bigg \}E^{2}\eta^{\alpha}.
\end{eqnarray}
Actually, this is {\it Ricci focusing} in $f(T,\mathcal{T})$ gravity as it's explained in
 the following. By make using $\eta^{\alpha}=\eta e^{\alpha}$, $e_{\alpha}e^{\alpha}=1$,
 $\epsilon_{\alpha}u^{\alpha}=e_{\alpha}k^{\alpha}=0$ and also writing an aligned base parallel
 propagated $\frac{De^{\alpha}}{D\nu}=k^{\beta}\nabla_{\beta}e^{\alpha}=0$, we get a new form of the null GDE (\ref{salako33}) as follows

\begin{eqnarray}\label{salako42}
{\frac{d^{2}\eta}{d\nu^{2}}=
\frac{-4\pi\,G}{(1+f_T)}\left(1+\frac{f_{\mathcal{T}}}{16\pi G}\right)
\Bigg\{ (\rho +\rho_1+p+p_1)  \Bigg \}E^{2}\eta^{\alpha}.}
\end{eqnarray}
At this  stage, the usual GR result discussed in
  \cite{2} can be recovered if we have
  $\kappa(\rho+p)>0$. 
  Hence, in a specific case  with the equation  of state $p=-\rho$
 (cosmological constant) the null geodesics notion is not affected.
  The relation  (\ref{salako42}) shows clearly that the 
 focusing condition for $f(T,\mathcal{T})$ gravity model
 provided that 
\begin{eqnarray}\label{salako43}
&&\frac{4\pi\,G}{(1+f_T)}\left(1+\frac{f_{\mathcal{T}}}{16\pi G}\right)
\Bigg\{ (\rho +\rho_1+p+p_1)  \Bigg \}> 0\cr
&& \frac{4 f_{TT} \dot{H} T + (f_{\mathcal{T}} + 16 G \pi) (p + \rho) + 
 2H f_{T\mathcal{T}} (-3 \dot{p}  + \dot{\rho})}{2 (1 + fT)}>0
\end{eqnarray}
Now, we can write the relation   (\ref{salako42}) in terms of redshift parameter $z$. Then we have 
\begin{eqnarray}\label{salako44}
\frac{d}{d\nu}=\frac{dz}{d\nu}\frac{d}{dz},
\end{eqnarray}

 which leads to
\begin{eqnarray}\label{salako45}
\frac{d^{2}}{d\nu^{2}}=\left(\frac{d\nu}{dz}\right)^{-2}\left[-\left(\frac{d\nu}{dz}\right)^{-1}\frac{d^{2}\nu}{dz^{2}}
\frac{d}{dz}+\frac{d^{2}}{dz^{2}}\right].
\end{eqnarray}
Let's  assume  the null geodesics governed  by
\begin{eqnarray}\label{salako46}
(1+z)=\frac{a_{0}}{a}=\frac{E}{E_{0}}\rightarrow\frac{dz}{1+z}=-\frac{da}{a}.
\end{eqnarray}
 Taking $a_{0}=1$ (the scale factor's current value), we obtain the following result for the past-directed case
\begin{eqnarray}\label{salako47}
dz=(1+z)\frac{1}{a}\frac{da}{d\nu}d\nu=(1+z)\frac{\dot{a}}{a}Ed\nu=E_{0}H(1+z)^{2}d\nu.
\end{eqnarray}
 Thus, we get
\begin{eqnarray}\label{salako48}
\frac{d\nu}{dz}=\frac{1}{E_{0}H(1+z)^{2}},
\end{eqnarray}
 and so
\begin{eqnarray}\label{salako49}
\frac{d^{2}\nu}{dz^{2}}=-\frac{1}{E_{0}H(1+z)^{3}}\left[\frac{1}{H}(1+z)\frac{dH}{dz}+2\right],
\end{eqnarray}
 where
\begin{eqnarray}\label{salako50}
\frac{dH}{dz}=\frac{d\nu}{dz}\frac{dt}{d\nu}\frac{dH}{dt}=-\frac{1}{H(1+z)}\frac{dH}{dt}.
\end{eqnarray}
 We recall here an important relation used in these previous equations: $\frac{dt}{d\nu}=E=E_{0}(1+z)$. From Hubble parameter's definition, 
 we derive the following relation
\begin{eqnarray}\label{salako51}
\dot{H}=\frac{\ddot{a}}{a}-H^{2}.
\end{eqnarray}
 Considering (\ref{salako38}), $\dot{H}$ becomes
\begin{eqnarray}\label{salako52}
\dot{H}=  \frac{-4\pi\,G}{(1+f_T)}\left(1+\frac{f_{\mathcal{T}}}{16\pi G}\right)
\Bigg\{\frac{\rho+\rho_1}{3} +(p+p_1)  \Bigg \}-H^{2},
\end{eqnarray}
where
\begin{eqnarray}\label{salako52}
-\frac{\dot{H}}{H^2}+2 =  \frac{4\pi\,G}{H^2(1+f_T)}\left(1+\frac{f_{\mathcal{T}}}{16\pi G}\right)
\Bigg\{\frac{\rho+\rho_1}{3} +(p+p_1)  \Bigg \}+3,
\end{eqnarray}
 thus, 
\begin{eqnarray}\label{salako53}
\frac{d^{2}\nu}{dz^{2}}=-\frac{3}{E_{0}H(1+z)^{3}}\left[  \frac{4\pi\,G}{3H^2(1+f_T)}\left(1+\frac{f_{\mathcal{T}}}{16\pi G}\right)
\Bigg(  \frac{\rho+\rho_1}{3} +(p+p_1)  \Bigg) + 1    \right].
\end{eqnarray}
Combining  this latter with   (\ref{salako45}), one obtains 
\begin{eqnarray}\label{salako54}
\frac{d^{2}\eta}{d\nu^{2}}=E_{0}H(1+z)^{2}\Bigg\{\frac{d^{2}\eta}{dz^{2}}+
\frac{3}{(1+z)}   \left[  \frac{4\pi\,G}{3H^2(1+f_T)}\left(1+\frac{f_{\mathcal{T}}}{16\pi G}\right)
\Bigg(  \frac{\rho+\rho_1}{3} +(p+p_1)  \Bigg) + 1    \right]     \frac{d\eta}{dz}\Bigg\}.
\end{eqnarray}
 Finally, by make using (\ref{salako42}), the null GDE takes the following form
\begin{eqnarray}\label{salako55}
&&\frac{d^{2}\eta}{dz^{2}}+
\frac{3}{(1+z)}   \left[  \frac{4\pi\,G}{3H^2(1+f_T)}\left(1+\frac{f_{\mathcal{T}}}{16\pi G}\right)
\Bigg(  \frac{\rho+\rho_1}{3} +(p+p_1)  \Bigg) + 1    \right]     \frac{d\eta}{dz}\cr
&&
-
\frac{4\pi\,G}{H^{2}(1+z)^{2}(1+f_T)}\left(1+\frac{f_{\mathcal{T}}}{16\pi G}\right)
\Bigg (\rho +\rho_1+p+p_1\Bigg )\eta=0.
\end{eqnarray}
 The contributions of matter and radiation to barotropic parameters $ \rho $ and $ p $ are written respectively as

\begin{eqnarray}\label{salako56}
\kappa \rho=3H_{0}^{2}\Omega_{m0}(1+z)^{3}+3H_{0}^{2}\Omega_{r0}(1+z)^{4},\hspace{20mm}
\kappa p=H_{0}^{2}\Omega_{r0}(1+z)^{4},
\end{eqnarray}
 where the following considerations $p_{m}=0$ and $p_{r}=\frac{1}{3}\rho_{r}$ have been done. By considering the equations in (\ref{salako56}), 
 the null GDE equation (\ref{salako55}) becomes
 \begin{eqnarray}\label{salako57}
\frac{d^{2}\eta}{dz^{2}}+P(H,\dot{H},z)\frac{d\eta}{dz}+Q(H,\dot{H},z)\eta=0,
\end{eqnarray}
where
\begin{eqnarray}\label{salako58}
 &&P(H,\dot{H},z) =  \frac{G\pi(1+\frac{f_{\mathcal{T}}}{16\pi G})}{3 (1 + f_T) H_0^2 (1 + z) 
 \Bigg(\Omega_{DE} + (1 + z)^3 \Big(\Omega_{m0} + \Omega_{r0} +  z \Omega_{r0}\Big)\Bigg)}\cr
 &&
 \Bigg\{    
\frac{(3 H_0^2 (1 + z)^4 \Omega_{r0}) }{  \pi G }
 +  \frac{ (6 H_0^2 (1 + z)^3 (\Omega_{m0} + \Omega_{r0} + z \Omega_{r0})) }{ \pi G } \cr
 &&
 - \frac{   (8 (-(f/2) + (f_{\mathcal{T}} H_0^2 (1 + z)^4 \Omega_{r0})/(8 \pi G) + 3 H_0^2 (\Omega_{DE} + (1 + z)^3 (\Omega_{m0} + \Omega_{r0} + 
  z \Omega_{r0}))))  }{ (f_{\mathcal{T}} + 16 G \pi) }
  \cr
  &&-      
 \frac{     (48 H (1 + z) ((-3 \frac{dp}{dz} + \frac{d\rho}{dz}) f_{T\mathcal{T}} H - 
   12 \frac{dH}{dz} f_{TT} H_0^2 (\Omega_{DE} + (1 + z)^3 (\Omega_{m0} + \Omega_{r0} + z \Omega_{r0}))))  }{ (f_{\mathcal{T}} + 16 G \pi)  }
\Bigg\},
\end{eqnarray}
\begin{eqnarray}\label{salako59}
 &&Q(H,\dot{H},z) =-  \frac{G\pi(1+\frac{f_{\mathcal{T}}}{16\pi G})}{ (1 + f_T) H_0^2 (1 + z)^2 
 \Bigg(\Omega_{DE} + (1 + z)^3 \Big(\Omega_{m0} + \Omega_{r0} +  z \Omega_{r0}\Big)\Bigg)}\cr
 &&
 \Bigg\{ \frac{( H_0^2 (1 + z)^4  \Omega_{r0}) }{ \pi G }
 +  \frac{ 3 H_0^2 (1 + z)^3 (\Omega_{m0} + \Omega_{r0} + z \Omega_{r0}) }{ \pi G } \cr
 &&-    
   \frac{  16 H (1 + z) ((-3 \frac{dp}{dz} + \frac{d\rho}{dz}) f_{T\mathcal{T}} H - 
   12 \frac{dH}{dz} f_{TT} H_0^2 (\Omega_{DE} + (1 + 
         z)^3 (\Omega_{m0} + \Omega_{r0} + 
         z \Omega_{r0}))) }{ (f_{\mathcal{T}} + 16 G \pi) }   
 \Bigg\},
\end{eqnarray}
in which  we have used the following new form of (\ref{salako39})
\begin{eqnarray}\label{salako60}
H^{2}=H_{0}^{2}[\Omega_{m0}(1+z)^{3}+\Omega_{r0}(1+z)^{4}+\Omega_{DE}],
\end{eqnarray}
 where $\Omega_{DE}$ has been defined as
\begin{eqnarray}\label{salako61}
\Omega_{DE}=-\frac{1}{6H_{0}^{2}} \Bigg[     
\frac{\left(f+12 H^{2} f_{T}\right)}{6}  + \frac{ f_{\mathcal{T}} (\rho+p)}{3}\Bigg].
\end{eqnarray}
To solve  Eq.(\ref{salako57}),  we have used  Eq.(\ref{salako30}).
Now, we will check the consistency of the found results  with those of the GR by choosing the special case
$ f (T, \mathcal {T}) = - 2 \Lambda $.
From this model, one obtains   $f_{T}=0$, $f_{\mathcal{T}}=0$ et  $f_{TT}=0$. So, $\Omega_{DE}$ in Eq.(\ref{salako61}) can be reduced to 

\begin{eqnarray}\label{salako62}
\Omega_{DE}=-\frac{1}{H_{0}^{2}} \Bigg[     
\frac{\left(-2\Lambda+12 H^{2} \times 0 \right)}{6}  + \frac{0 \times  (\rho+p)}{3}\Bigg]=\frac{\Lambda}{3H_{0}^{2}}\equiv\Omega_{\Lambda}.
\end{eqnarray}
This relation allows us to rewrite the first Friedmann equation in GR in the  following form 
\begin{eqnarray}\label{salako63}
H^{2}=H_{0}^{2}[\Omega_{m0}(1+z)^{3}+\Omega_{r0}(1+z)^{4}+\Omega_{\Lambda}].
\end{eqnarray}
  
 Thus,  $P$ and $Q$ become dependant on  only the redshift parameter  as 
\begin{eqnarray}\label{salako64}
P(z)=\frac{\frac{7}{2}\Omega_{m0}(1+z)^{3}+4\Omega_{r0}(1+z)^{4}+2\Omega_{\Lambda}}{(1+z)[\Omega_{m0}(1+z)^{3}+\Omega_{r0}(1+z)^{4}+\Omega_{\Lambda}]},
\end{eqnarray}

\begin{eqnarray}\label{salako65}
Q(z)=\frac{3\Omega_{m0}(1+z)+4\Omega_{r0}(1+z)^{2}}{2[\Omega_{m0}(1+z)^{3}+\Omega_{r0}(1+z)^{4}+\Omega_{\Lambda}]}.
\end{eqnarray}
 Ultimately, the GDE for null vector fields becomes
\begin{eqnarray}\label{salako66}
\frac{d^{2}\eta}{dz^{2}}+\frac{\frac{7}{2}\Omega_{m0}(1+z)^{3}+4\Omega_{r0}(1+z)^{4}+
2\Omega_{\Lambda}}{(1+z)[\Omega_{m0}(1+z)^{3}+\Omega_{r0}(1+z)^{4}+\Omega_{\Lambda}]}\frac{d\eta}{dz}+
\frac{3\Omega_{m0}(1+z)+4\Omega_{r0}(1+z)^{2}}{2(\Omega_{m0}(1+z)^{3}+\Omega_{r0}(1+z)^{4}+\Omega_{\Lambda})}\eta=0.
\end{eqnarray}
 We emphasize here that in order to obtain the Mattig relation in GR \cite{sch}, 
 we have to fix $\Omega_{\Lambda}=0$, $\Omega_{r0}+\Omega_{m0}=1$ which leads to

\begin{eqnarray}\label{salako67}
\frac{d^{2}\eta}{dz^{2}}+\frac{\frac{7}{2}\Omega_{m0}(1+z)^{3}+4\Omega_{r0}(1+z)^{4}}{(1+z)[\Omega_{m0}(1+z)^{3}+
\Omega_{r0}(1+z)^{4}]}\frac{d\eta}{dz}+\frac{3\Omega_{m0}(1+z)+4\Omega_{r0}(1+z)^{2}}{2(\Omega_{m0}(1+z)^{3}+\Omega_{r0}(1+z)^{4})}\eta=0.
\end{eqnarray}
Then we can use (\ref{salako57}) to generalize  Mattig relation in  $ f (T,  \mathcal {T}) $ gravity.  These previous results
can be used to generate the observer area distance $r_{0}(z)$ \cite{sch}

\begin{eqnarray}\label{salako68}
r_{0}(z)=\sqrt{\left|\frac{dA_{0}(z)}{d\Omega}\right|}=\left|\frac{\eta(z')|_{z}}{d\eta(z')/d\ell|_{z'=0}}\right|,
\end{eqnarray}
where $A_{0}$ is the area of the object and also $\Omega$ is the solid angle. 
Having  $d/d\ell=E^{-1}_{0}(1+z)^{-1}d/d\nu=H(1+z)d/dz$ and reducing to zero 
the deviation at $z=0$,  we consequently obtain

\begin{eqnarray}\label{salako69}
r_{0}(z)=\left|\frac{\eta(z)}{H(0) d\eta(z')/dz'|_{z'=0}}\right|.
\end{eqnarray}
$H(0)$ is the result of the modifified Friedmann equation evaluation at $z=0$.

 \subsection{ Numerically solution of GDE for null vector fields in $f(T,\mathcal{T})$ gravity}
 In order to solve numerically the null vector GDE  in $f(T,\mathcal{T})$ gravity, we have considered the  model
 of $f(T,\mathcal{T})=T +f(\mathcal{T})$ gravity, where  $ f(\mathcal{T})= \gamma \mathcal{T}^{\sigma}$; $\gamma$
 and $\sigma = \frac{1+3w}{2(1+w)}$ 
 being constant. 
 A thorough study of this cosmological model shows very quickly
   interesting results that can be found in  \cite{salakonew}. Considering this model,  the equations  
   (\ref{salako58}), (\ref{salako59}, (\ref{salako61}) can be rewritten under the following forms  
\begin{eqnarray}
\Omega_{DE} =
 \frac{  -\gamma \; \mathcal{T}^{-1 + \sigma} \Bigg(4 \pi G \mathcal{T} + 
    H_0^2 (1 + z)^3 \sigma (3 \Omega_{m0} + 
       4 (1 + z) \Omega_{r0})\Bigg )}{24 H_0^2 \pi G) },
\end{eqnarray}
\begin{eqnarray}
&&P(H,\frac{dH}{dz},z) =            
 \frac{(\gamma \mathcal{T}^\sigma (2 \pi G \mathcal{T} + 
    H_0^2 (1 + z)^3 \sigma (3 \Omega_{m0} + 
       4 (1 + z) \Omega_{r0})) }{(24 H_0^2 (1 + 
   z) \pi G \mathcal{T} (\Omega_{DE} + (1 + 
      z)^3 (\Omega_{m0} + \Omega_{r0} + 
      z \Omega_{r0}))) }\cr
      &&- \frac{
 12 H_0^2 \mathcal{T} (-2 G \pi (1 + z)^3 (2 \Omega_{m0} + 
       3 (1 + z) \Omega_{r0}) + \pi G (\Omega_{DE} + (1 +
           z)^3 (\Omega_{m0} + \Omega_{r0} + 
          z \Omega_{r0})))) }{(24 H_0^2 (1 + 
   z) \pi G \mathcal{T} (\Omega_{DE} + (1 + 
      z)^3 (\Omega_{m0} + \Omega_{r0} + 
      z \Omega_{r0})))}, 
\end{eqnarray}
\begin{eqnarray}
Q(H,\frac{dH}{dz},z) = \frac{ -(((1 + z) (16 G \pi \mathcal{T} + \gamma \sigma \mathcal{T}^\sigma) (3 \Omega_{m0} + 4 (1 + z)
\Omega_{r0}))}{(
 16 \pi G \mathcal{T} (\Omega_{DE} + (1 + 
       z)^3 (\Omega_{m0} + \Omega_{r0} + 
       z \Omega_{r0}))))},
\end{eqnarray}
with
\begin{eqnarray}
\mathcal{T}=    
      \frac{  3 H_0^2 (1 + z)^3 \Omega_{m0}}{(8 \pi G)}.
\end{eqnarray}

 \begin{figure}[h]
  \centering
  \begin{tabular}{rl}
\includegraphics[width=9cm, height=7cm]{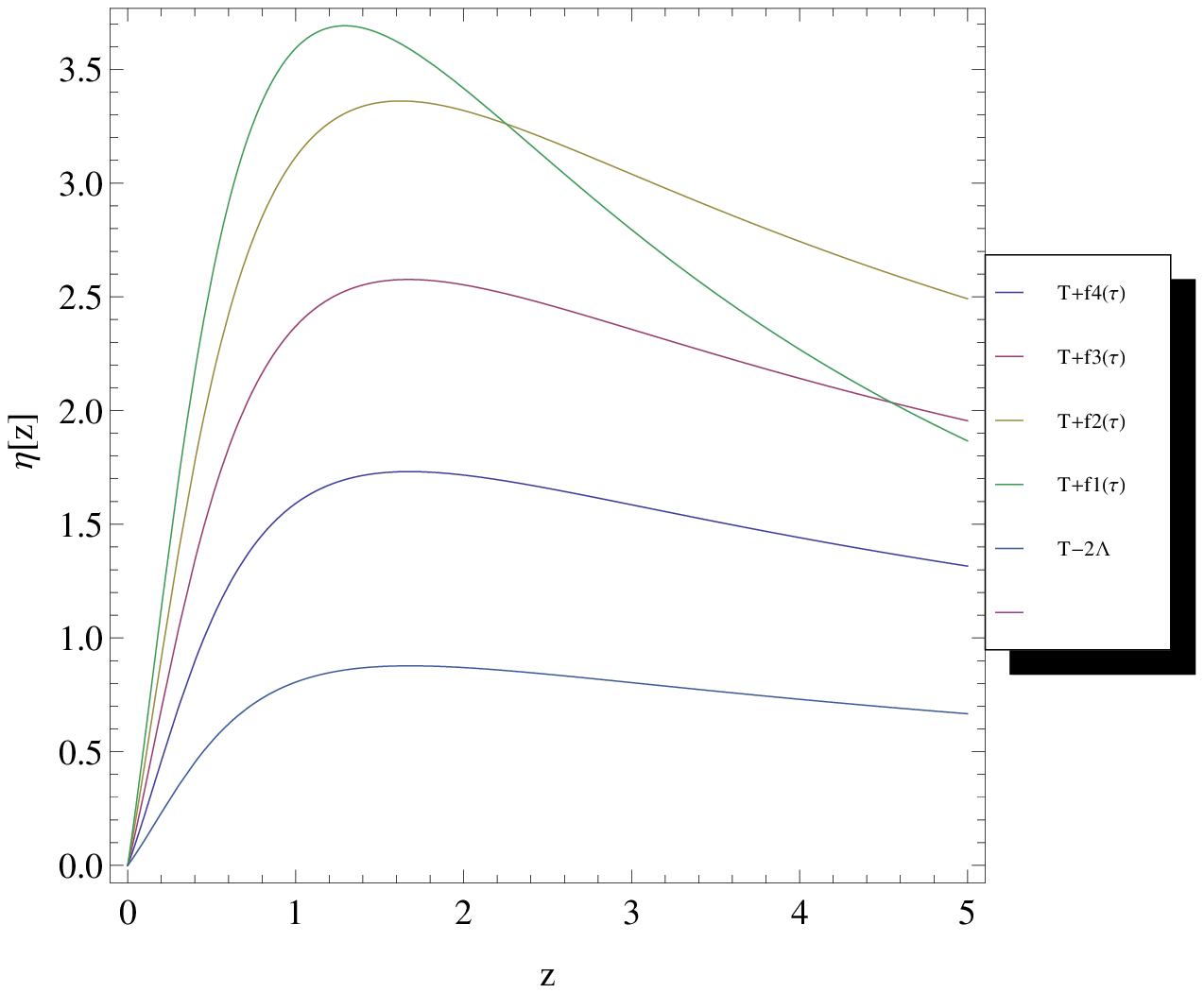}&
 \includegraphics[width=9cm, height=7cm]{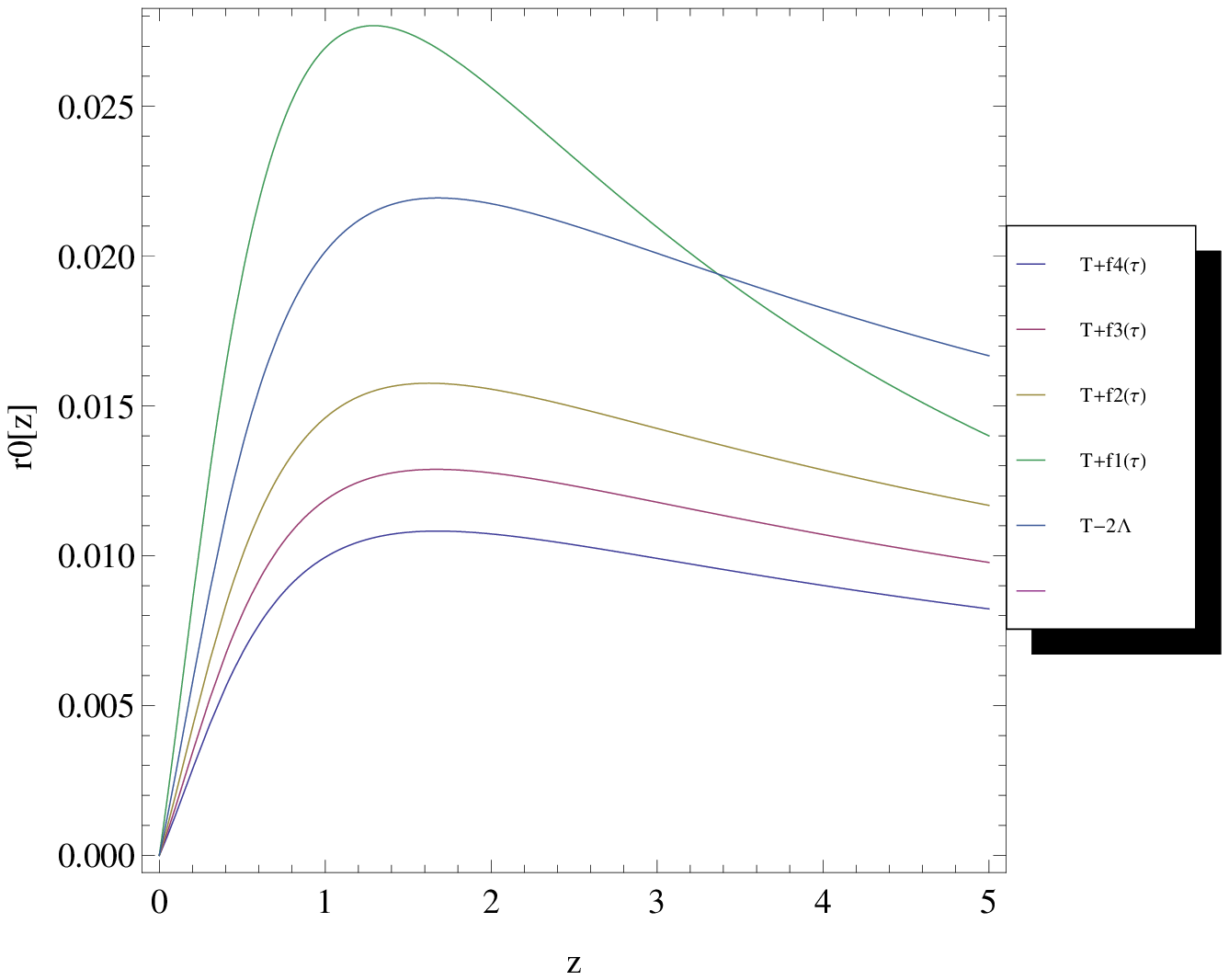}
  \end{tabular}
\caption{The graphs shows the deviation vector magnitude $\eta(z)$ (left panel) and observer area distance $r_0(z)$ (right panel) for null
  vector field GDE with FLRW background as functions of redshift. The graphs are plotted for
   $H_0 = 80Km/s/Mpc$, $\Omega_{m0}=0.3$, $\Omega_{r0}=\Omega_{k0}=0$, $\Lambda=1.7.10^{-121}$ 
  and we imposed in equation (\ref{salako57}) the initial conditions $\eta(z=0)= 0$ and $\eta'(z=0)=1$.}
 \label{fig2}
 \end{figure} 
In each panel of Figure 2, we observe that the curves reflecting the evolution of the intensity of deviation vector $ \eta(z) $ 
and the distance of the area of the observer $r_0(z)$ have similar behaviors to that of the model $ \Lambda CDM $.
Within the
Model $ f (T, \mathcal {T}) = \gamma T^ {\sigma} $, we see that when $ \gamma $ is increasing ($\gamma \geqslant 1$)  and we are going to
larger values of  redshift $(z \geqslant 0.8) $, the deviation vector  intensity  $ \eta $ (z) and
observer area distance  $r_0(z)$  decouple from each model $ \Lambda CDM $ 
but still keep the same pace while for small values of redshifts namely nowadays
the model $ f (T, \mathcal {T}) $ can accurately replicate the model $ \Lambda CDM $. 
We can then conclude that for all  considered cases , the results are compactible to $ \Lambda CDM $. So  the above studied
 $f(T,\mathcal{T})$  models  remain phenomenologically viable and can be tested with observational data.

 \section{Conclusion}\label{sec5}

 In this paper, we have presented the Geodesic Deviation Equation (GDE) in the context of $f(T,\mathcal{T})$ gravity applied to metric. The
 determination by rigorous calculation of the scalar Ricci and Riemann tensor was firstly executed by  using the 
$ f(T, \mathcal {T}) $ gravity  field equations. The Geodesic Deviation Equation and the generalization of {\it Pirani } equation for the FLRW Universe in $f(T,\mathcal{T})$
 gravity have been investigated and both these equations have been reduced to the well known Mattig relation when $f(T,\mathcal{T}) = -2\Lambda$. 
 We have performed for two particular cases, the GDE for fundamental observes and the past-directed null vector fields with FLRW Universe.
  Within these cases we have obtained the Raychaudhuri equation, the generalized Mattig relation and the diametric angular distance 
  differential for $f(T,\mathcal{T})$ gravity theory.  Furthermore, as it is usually done in  GR,  we have also investigated
  the past-directed null geodesics condition for $f(T,\mathcal{T})$ gravity.
  Numerical results concerning the geodesic deviation $ \eta (z) $ and the observer area distance $r_0(z)$ 
  for $f (T, \mathcal {T})$ models were found and compared with their equivalent supplied by the $\Lambda CDM$  model.

\section*{Acknowledgments}
 The authors  thank IMSP for hospitality during the elaboration of this work.


\begin{thebibliography}{9}

\bibitem{Misner}C. W. Misner, K. S. Thorne, and J. H. Wheeler, {\it Gravitation}, W. H. Freeman and Company, 1973.
\bibitem{eti25} Szekeres P: The Gravitational Compass, J. Maths. Phys. 6 (1965), 1387.
\bibitem{eti20} J. L. Synge. On the Deviation of Geodesics and Null Geodesics, Particularly in Relation to the
Properties of Spaces of Constant Curvature and Indefinite Line Element. Ann. Math. 35:705 (1934).
\bibitem{eti21}
F. A. E. Pirani. On the Physical Significance of the Riemann Tensor. Acta Phys. Polon. 15:389
(1956).
\bibitem{eti22} G. F. R. Ellis and H. Van Elst. Deviation of geodesics in FLRW spacetime geometries (1997). Preprint
in [arXiv:gr-qc/9709060v1].
\bibitem{eti23} S.L. Shapiro and S.A. Teukolsky, Black Holes, White Dwarfs and Neutron Satrs (Wile-Interscience, New York 1983).
\bibitem{eti24} K.S. Thorne, in S. Hawking and W. Israel, eds, 300 Year of
Gravitation (Cambridge University, Cambridge 1987) p. 330.
\bibitem{eti26} Raychaudhuri A K: Relativistic Cosmology, Phys. Rev. 98 (1955), 1123.
\bibitem{eti27} Mattig W: Uber den Zusammenhang zwischen Rotverschiebung und scheinbarer Helligkeit, Astr.
Nach. 284 (1958), 109.
\bibitem{eti28} Pirani F A E: On the Physical Significance of the Riemann Tensor, Acta Phys. Polon. 15 (1956),389.
\bibitem{Spergel} D. N. Spergel, et. al., Astrophys. J. Suppl. {\bf 170}, (2007)  377, arXiv:astro-ph/0603449.
\bibitem{Adelman} J. K. Adelman-McCarthy, et. al., Astrophys. J. Suppl. {\bf 175}, (2008) 297, arXiv:0707.3413[astro-ph].
\bibitem{a'}
R.Aldrovandi and J.G.Pereira,TELEPARALLEL GRAVITY, \\in
http://www.ift.unesp.br/users/jpereira/tele.pdf.
\bibitem{ma1}{\bf A. De Felice and S. Tsujikawa, Living Rel. Rev. {\bf 13}, 3 (2010) [arXiv:1002.4928
[gr-qc]]; K. Bamba, S. Capozziello, S. Nojiri, S. D. Odintsov.  Astrophys. Space Sci. {\bf 342}, 155 (2012) [arXiv:1205.3421 [gr-qc]]};
S. Nojiri and S. D. Odintsov,  	ECONF C {\bf 0602061}, 06 (2006); Int. J. Geom. Meth. Mod. Phys. {\bf 4}, 115-146 (2007) [arXiv:hep-th/0601213];   Phys. Rept. {\bf 505}, 59-144 (2011) [arXiv:1011.0544].
\bibitem{premierfrt}
T. Harko, F. S. N. Lobo, S. Nojiri and S. D. Odintsov, ?f(R, T ) gravity,? Phys. Rev. D {\bf 84} (2011) 024020. [arXiv:1104.2669 [gr-qc]].
\bibitem{ma2}
 M. J. S. Houndjo, Int. J. Mod. Phys. D. {\bf 21}, 1250003 (2012). arXiv: 1107.3887 [astro-ph.CO].
\bibitem{ma3} 
 M. J. S. Houndjo and O. F. Piattella, Int. J. Mod. Phys. D. {\bf 21}, 1250024 (2012). arXiv: 1111.4275 [gr.qc].
 \bibitem{ma4}
  D. Momeni, M. Jamil and R. Myrzakulov, Euro. Phys. J. C {\bf 72}, arXiv: 1107.5807[physics.gen-ph]. 
\bibitem{ma5}  
 M. J. S. Houndjo, C. E. M. Batista, J. P. Campos and O. F. Piattella,?? [arXiv:1203.6084 [gr-qc]].
\bibitem{ma6}
F. G. Alvarenga, M. J. S. Houndjo, A. V. Monwanou and Jean. B. Chabi-Orou,?? arXiv: 1205.4678 [gr-qc].
\bibitem{mj1}  S.~'i.~Nojiri and S.~D.~Odintsov,
   ``Modified Gauss-Bonnet theory as gravitational alternative for dark
energy,''
   Phys.\ Lett.\ B {\bf 631}, 1 (2005)  [hep-th/0508049]; 
S. Nojiri, S. D. Odintsov, A. Toporensky, P. Tretyakov, arXiv:0912.2488.
\bibitem{mj2}
K. Bamba, S. D. Odintsov, L. Sebastiani, S. Zerbini, arXiv:0911.4390. 
\bibitem{mj3}
K. Bamba, C.-Q. Geng, S. Nojiri, S. D. Odintsov, arXiv:0909.4397.
\bibitem{mj4}
M.E. Rodrigues, M.J.S. Houndjo, D. Momeni, R. Myrzakulov, arXiv:1212.4488.
\bibitem{mj5}
M. J. S. Houndjo, M. E. Rodrigues, D. Momeni, R. Myrzakulov .
 arXiv:1301.4642 [gr-qc].
\bibitem{st1} J.~Amor\'os, J.~de Haro and S.~D.~Odintsov,
   ``Bouncing Loop Quantum Cosmology from $f(T)$ gravity,''
   Physical Review D 87, {\bf 104037} (2013)
   [arXiv:1305.2344 [gr-qc]]; 
  K.~Bamba, J.~de Haro and S.~D.~Odintsov,
   ``Future Singularities and Teleparallelism in Loop Quantum Cosmology,''
   JCAP {\bf 1302} (2013) 008
   [arXiv:1211.2968 [gr-qc]];
  K.~Bamba, S.~'i.~Nojiri and S.~D.~Odintsov,
   ``Effective $f(T)$ gravity from the higher-dimensional Kaluza-Klein and
Randall-Sundrum theories,''
   arXiv:1304.6191 [gr-qc]; 
G. R. Bengochea, R. Ferraro and , Phys. Rev. D {\bf 79}, 124019 (2009) [arXiv:0812.1205 [astro-ph]].
\bibitem{st2}
 E. V. Linder, Phys.Rev. D {\bf 81}, 127301 (2010) [Erratum-ibid. D 82, 109902 (2010)] [arXiv:1005.3039 [astro-ph.CO]].
\bibitem{st3}
 M. Jamil, D. Momeni and R. Myrzakulov, Eur. Phys. J. C {\bf 72} (2012) 2267 [arXiv:1212.6017 [gr-qc]].
\bibitem{st4} 
  R. Myrzakulov, Entropy {\bf 14} (2012) 1627[arXiv:1212.2155 [gr-qc]]. 
\bibitem{st5}  
   M. R. Setare and N. Mohammadipour, JCAP {\bf 1211} (2012) 030 [arXiv:1211.1375 [gr-qc]].
\bibitem{st6}   
    M.R. Setare, N. Mohammadipour, JCAP {\bf 01} (2013) 015 [arXiv: 1301.4891].
\bibitem{st7}   
 M. Jamil, D. Momeni, R. Myrzakulov and P. Rudra, J. Phys. Soc. Jap. {\bf 81}  (2012) 114004 [arXiv:1211.0018 [physics.gen-ph]].
 \bibitem{st8}
  M. E. Rodrigues, M. J. S. Houndjo, D. Saez-Gomez and F. Rahaman, Phys. Rev. D {\bf  86} (2012) 104059 [arXiv:1209.4859 [gr-qc]].
 \bibitem{st9} 
 M. Jamil, D. Momeni and R. Myrzakulov, Eur. Phys. J. C {\bf 72} (2012) 2122 [arXiv:1209.1298 [gr-qc]].
 \bibitem{st10}
  R. Myrzakulov, Eur. Phys. J. C {\bf 72} (2012)2203 [arXiv:1207.1039 [gr-qc]].
  \bibitem{st11}
   M. J. S. Houndjo, D. Momeni and R. Myrzakulov, Int. J. Mod. Phys. D {\bf 21} (2012)
1250093 [arXiv:1206.3938 [physics.gen-ph]].
\bibitem{st12}
 M. E. Rodrigues, M. H. Daouda and M. J. S. Houndjo, arXiv:1205.0565
[gr-qc].
\bibitem{st13}
 M. R. Setare and M. J. S. Houndjo, arXiv:1203.1315 [gr-qc].
 \bibitem{st14}
  K. Bamba, M. Jamil, D. Momeni and R. Myrzakulov, arXiv:1202.6114 [physics.gen-ph].
 \bibitem{st15} 
 K. Bamba, R. Myrzakulov, S. 'i. Nojiri and S. D. Odintsov, Phys. Rev. D {\bf 85} (2012)104036 [arXiv:1202.4057 [gr-qc]].
 \bibitem{st16}
  M. Jamil, D. Momeni and R. Myrzakulov, Eur. Phys. J. C {\bf 72} (2012)  2267 [arXiv:1212.6017[gr-qc]].
  \bibitem{st17}
   M. Jamil, D. Momeni and R. Myrzakulov, Gen. Rel. Grav. {\bf 45} (2013) 263 [arXiv:1211.3740 [physics.gen-ph]].
   \bibitem{st18}
M. Jamil, D. Momeni and R. Myrzakulov, Eur. Phys. J. C {\bf 72} (2012) 2122 [arXiv:1209.1298 [gr-qc]].
\bibitem{st20}
 M. Jamil, D. Momeni and R. Myrzakulov, Eur. Phys. J. C {\bf 72} (2012) 2075 [arXiv:1208.0025 [gr-qc]].
 \bibitem{st21}
  M. Jamil, K. Yesmakhanova, D. Momeni and R. Myrzakulov, Central Eur. J. Phys. {\bf 10} (2012) 1065 [arXiv:1207.2735 [gr-qc]].
  \bibitem{st22}
   M. J. S. Houndjo, D. Momeni and R. Myrzakulov, Int. J. Mod. Phys. D {\bf 21} (2012) 1250093 [arXiv:1206.3938 [physics.gen-ph]].
   \bibitem{st23}
 M. Jamil, D. Momeni and R. Myrzakulov, Eur. Phys. J. C {\bf 72} (2012) 1959 [arXiv:1202.4926 [physics.gen-ph]].
 \bibitem{st24}
  M. H. Daouda, M. E. Rodrigues and M. J. S. Houndjo, Phys. Lett. B {\bf 715} (2012) 241 [arXiv:1202.1147 [gr-qc]].
 \bibitem{st25} 
 M. Jamil, S. Ali, D. Momeni, R. Myrzakulov and Eur. Phys. J. C {\bf  72}, 1998 (2012) [arXiv:1201.0895 [physics.gen-ph]].
 \bibitem{st26}
  M. Jamil, D. Momeni, N. S. Serikbayev, R. Myrzakulov and , Astrophys. Space Sci.{ \bf 339}, 37 (2012) [arXiv:1112.4472 [physics.gen-ph]].
  \bibitem{st27}
   M. Jamil, D. Momeni, M. A. Rashid and , Eur. Phys. J. C {\bf  71}, 1711 (2011) [arXiv:1107.1558 [physics.gen-ph]].
   \bibitem{st28}
 M. Hamani Daouda, M. E. Rodrigues and M. J. S. Houndjo, Eur. Phys. J. C {\bf 72} (2012) 1893 [arXiv:1111.6575 [gr-qc]].
 \bibitem{st29}
  M. Hamani Daouda, M. E. Rodrigues and M. J. S. Houndjo, Eur. Phys. J. C {\bf 72} (2012) 1890 [arXiv:1109.0528 [physics.gen-ph]].
  \bibitem{st30}
   R. Myrzakulov, Gen. Rel. Grav. {\bf 44} (2012) 3059 [arXiv:1008.4486 [physics.gen-ph]].
   \bibitem{st31}
 K. K. Yerzhanov, S. .R. Myrzakul, I. I. Kulnazarov and R. Myrzakulov, arXiv:1006.3879 [gr-qc].
 \bibitem{st32}
  R. Myrzakulov, Eur. Phys. J. C {\bf 71} (2011) 1752 [arXiv:1006.1120 [gr-qc]].
  \bibitem{st33}
  M. E. Rodrigues, M. J. S. Houndjo, D. Momeni, R. Myrzakulov and , arXiv:1302.4372 [physics.gen-ph].
\bibitem{st34}
 J. M. Bardeen, B. Carter, S. W. Hawking , Commun. Math. Phys. {\bf 31} (1973) 161-170.
 \bibitem{st35}
 N. Tamanini and C. G. Boehmer, Phys. Rev. D {\bf 86}, 044009 (2012), arXiv:1204.4593 [gr-qc].
 \bibitem{st36}
 Baojiu Li, T. P. Sotiriou and J. D. Barrow, Phys. Rev. D {\bf 83}, 064035 (2011); Phys. Rev. D {\bf 83}, 104030 (2011).
 \bibitem{st37}
 M. J. S. Houndjo, D. Momeni, R. Myrzakulov and M. E. Rodrigues,arXiv:1304.1147.
\bibitem{st38}
 C. Deliduman and B. Yapiskan, arXiv:1103.2225v3 [gr-qc].
\bibitem{st39}
 M. Hamani Daouda, M. E. Rodrigues and M. J. S. Houndjo, Eur. Phys. J. C {\bf 71} (2011) 1817 [arXiv:1108.2920 [astro-ph.CO]].
 \bibitem{st40} 
 I.G.Salako, M.E.Rodrigues, A.V.Kpadonou, M.
J.S.Houndjo and J.Tossa
 JCAP { \bf 060}, 1475-7516 (2013) 
 \bibitem{st41} 
M. E. Rodrigues, I. G. Salako, M. J. S. Houndjo, J. Tossa Int. J. Mod. Phys. D { \bf 23}, 1450004 (2014) 
 \bibitem{sala1}
 Davood Momeni, Ratbay Myrzakulov.: arXiv:1405.5863 [gr-qc] 
 \bibitem{sala2} 
Tiberiu Harko, Francisco S. N. Lobo, G. Otalora,and Emmanuel N. Saridakis http://arxiv.org/abs/1405.0519v1
\bibitem{sala3} 
Ines G. Salako, Abdul Jawad, Surajit Chattopadhyay, Astrophys.Space Sci. 358 (2015) 1, 13

\bibitem{sala4} 
S. B. Nassur, M. J. S. Houndjo, A. V. Kpadonou, M. E. Rodrigues, J. Tossa, http://arxiv.org/abs/1506.09161
\bibitem{Ferraro1}
 R. Ferraro and F. Fiorini, Phys. Rev. D {\bf 75} (2007) 084031
 \bibitem{Ferraro2}
G. R. Bengochea and R. Ferraro, Phys. Rev. D {\bf 79}(2009)124019 
\bibitem{eti35}
A. Guarnizo, L. Castaneda, J. M. Tejeiro, Gen. Rel. Grav. 43, 2713 (2011);
A. de la Cruz-Dombriz, P. K. S. Dunsby, V. C. Busti, S. Kandhai, Phys. Rev. D 89, 064029 (2014), arXiv:1312.2022;
A. Guarnizo, L. Castaneda, J. M. Tejeiro, arXiv:1402.3196.

\bibitem{eti36}
F. Shojai and A. Shojai, Phys. Rev. D 78, 104011 (2008).

\bibitem{eti37} F. Darabi, M. Mousavi, K. Atazadeh,  Dec 31, 2014. 11 pp.
Published in Phys.Rev. D91 (2015) 084023.
\bibitem{etienne}
E. H. Baffou, M. J. S. Houndjo, M. E. Rodrigues, A. V. Kpadonou, J. Tossa, arXiv:1509.06997 
\bibitem{3} R. M. Wald, {\it General Relativity}, The University of Chicago Press, 1984;\\
 E. Poisson, {\it A Relativist's Toolkit - The Mathematics of Black-Hole Mechanics}, Cambridge University
Press, 2004.
\bibitem{2} J. L. Synge, Ann. Math. {\bf35}, 705 (1934);\\
 F. A. E. Pirani, Acta Phys. Polon. {\bf15}, 389 (1956);\\ G. F. R. Ellis and H. Van Elst, [arXiv:gr-qc/9709060v1].
 \bibitem{ellis}G. F. R. Ellis and H. Van Elst, [arXiv:gr-qc/9812046v5].
 \bibitem{cac}D. L. Caceres, L. Casta~neda, J. M. Tejeiro, J.
Phys. Conf. Ser. {\bf229},  012076 (2010), [arXiv:0912.4220v1].

\bibitem{rafael2}G.R. Bengochea and R. Ferraro, Phys. Rev. D {\bf79},  124019 (2009), [arXiv:0812.1205].
\bibitem{sch}P. Schneider, J. Ehlers and E. E. Falco, {\it Gravitational Lenses}, (Springer-Verlag, 1999).
\bibitem{salakonew}
Ines.G.Salako,  M. J. S. Houndjo, M. E. Rodrigues, and A. V. Kpadonou, to Appear

\end{thebibliography}
\end{document}